\documentclass[prl,aps,twocolumn,showpacs,superscriptaddress]{revtex4-2}

\usepackage{graphicx}
\usepackage{graphicx, color}
\usepackage{hyperref}
\definecolor{link}{rgb}{0.1,0.1,0.9}
\hypersetup{colorlinks=true,linkcolor=link,citecolor=link,urlcolor=link,linktocpage}
\usepackage{dcolumn}
\usepackage{bm}
\usepackage{gensymb}
\usepackage[utf8]{inputenc}
\usepackage[T1]{fontenc}
\usepackage{mathptmx} 
\usepackage{booktabs}
\usepackage{amsmath}

\begin{document}

\title{Dirac topology, anomalous Hall response, and giant magnetoresistance in carrier-compensated altermagnetic semimetal NiS}

\author{Shovan Gayen} 
 \affiliation{Department of Physics, Bennett University, Greater Noida 201310, Uttar Pradesh, India}
 \author{Sk. Soyeb Ali}%
 \affiliation{Department of Physics, Bennett University, Greater Noida 201310, Uttar Pradesh, India}
\author{S K Panda}
\email[Corresponding author: ]{swarup.panda@bennett.edu.in}
\affiliation{Department of Physics, Bennett University, Greater Noida 201310, Uttar Pradesh, India}


\begin{abstract}
We combine first-principles density-functional theory, Berry-curvature analysis, semiclassical Boltzmann transport, and atomistic spin dynamics to establish hexagonal NiS as a compensated $3d$ altermagnetic semimetal in which topology, magnetism, and lattice dynamics are intrinsically intertwined. The rotational coset symmetry of the NiAs lattice produces momentum-dependent spin splitting characteristic of altermagnetism. With spin–orbit coupling, gapped Dirac-like crossings generate intense Berry-curvature hot spots and nearly compensated electron–hole pockets. This leads to a large and anisotropic intrinsic spin Hall conductivity comparable to that of several $4d$/$5d$ metals, a symmetry-allowed anomalous Hall response despite zero net magnetization, and nonsaturating magnetoresistance exceeding $10^3$\%.
On the magnetic side, first-principles determination of the exchange tensor reveals dominant long-range superexchange and sizable anisotropic interactions, quantitatively reproducing the experimental N\'eel temperature. Our results identify NiS as a model $3d$ platform in which carrier compensation, altermagnetic symmetry, Berry-curvature–driven transport, and lattice-sensitive magnetism coexist within a single symmetry framework, offering a design principle for multifunctional quantum responses in correlated transition-metal compounds.
\end{abstract}

\maketitle
The interplay between electronic topology, magnetism, and lattice dynamics underlies many of the most striking emergent phenomena in condensed-matter physics. Three-dimensional (3D) topological Dirac semimetals (DSMs) realize a quantum state in which low-energy excitations behave as massless Dirac fermions, effectively two degenerate Weyl fermions of opposite chirality~\cite{topology1,topology2,wang2013three}. Although a Dirac point carries zero net Chern number, its fourfold degeneracy can be stabilized by crystalline symmetries, including $C_{4v}$, $C_{6v}$, and nonsymmorphic operations~\cite{topology3,fang2016topological}. Such symmetry-protected Dirac and Weyl semimetals exhibit unconventional transport responses, most notably extremely large magnetoresistance (XMR), which is of both fundamental and technological interest~\cite{Ali_WTe2,shekhar2015extremely,tafti2016resistivity}. A key microscopic mechanism underlying XMR is nearly perfect electron-hole compensation, whereby balanced carrier populations lead to a dramatic suppression of conductivity under magnetic field~\cite{pippard1989magnetoresistance,Ali_WTe2}. When combined with symmetry-protected Dirac-like dispersions, this compensation yields a robust semimetallic state with enhanced transport responses~\cite{burkov2016topological,armitage2018weyl}. 
\par 
In the quest for a material uniting Dirac-like electronic structure, electron–hole compensation, anomalous Hall response, and large magnetoresistance , hexagonal nickel sulfide (NiS) emerges as a striking case study. Crystallizing in the hexagonal NiAs-type structure (space group $P6_3/mmc$)\cite{cryststruct}, NiS undergoes a first-order transition at $T_t \simeq 265$K from a high-temperature Pauli paramagnetic metal to a low-temperature A-type antiferromagnetic (AFM) semimetal\cite{sparks2,neutron,neutron1,photo1,photoemission,mypaper,anisimov}. In the antiferromagnetic phase, Ni moments order ferromagnetically within the $ab$ plane and antiferromagnetically along the $c$ axis~\cite{neutron1}, accompanied by an abrupt $\sim2\%$ volume expansion~\cite{sparks2}. Recent theoretical work~\cite{Altermagnet-NiS} has identified this phase as altermagnetic, where antiferromagnetic order coexists with momentum-dependent spin splitting despite zero net magnetization. Additionally, large magnetoresistance has been reported experimentally for this system~\cite{mr2}. These features motivate the exploration of NiS as a candidate altermagnetic Dirac semimetal where topology, and magnetism are intrinsically linked.
\par 
Here, we present a comprehensive investigation of the low-temperature altermagnetic phase of NiS, revealing the cooperative interplay of Dirac-like band topology, carrier compensation, and pronounced magnetotransport responses. Our analysis combines density functional theory (DFT) calculations with model-based approaches, and methodological details are provided in the Supplemental Material (SM)~\cite{SM}. Within this framework, we identify symmetry-protected Dirac-like band crossings, a strongly anisotropic intrinsic spin Hall conductivity (SHC), and nearly perfect electron–hole compensation that gives rise to giant magnetoresistance.
Intense Berry-curvature hot spots further generate a finite intrinsic anomalous Hall conductivity (AHC), demonstrating that altermagnetic symmetry lifts the conventional constraint forbidding Berry-curvature-driven charge transport in collinear $3d$ antiferromagnets. From first principles, we extract the complete magnetic interaction tensor, including isotropic exchange, Dzyaloshinskii-Moriya (DM) interactions, and single-ion anisotropy and employ stochastic Landau-Lifshitz-Gilbert (sLLG) simulations~\cite{SLLG1,SLLG2} to capture the temperature-dependent magnetic behavior. Taken together, these results resolve long-standing magnetic and transport anomalies in NiS and establish it as a prototypical altermagnetic compensated Dirac semimetal hosting large intrinsic spin and anomalous Hall responses, and giant magnetoresistance, pointing to general design principles for multifunctional quantum materials where topology, magnetism, and lattice symmetry are tightly intertwined.
\begin{figure}[t]
\includegraphics[width=1.0\columnwidth]{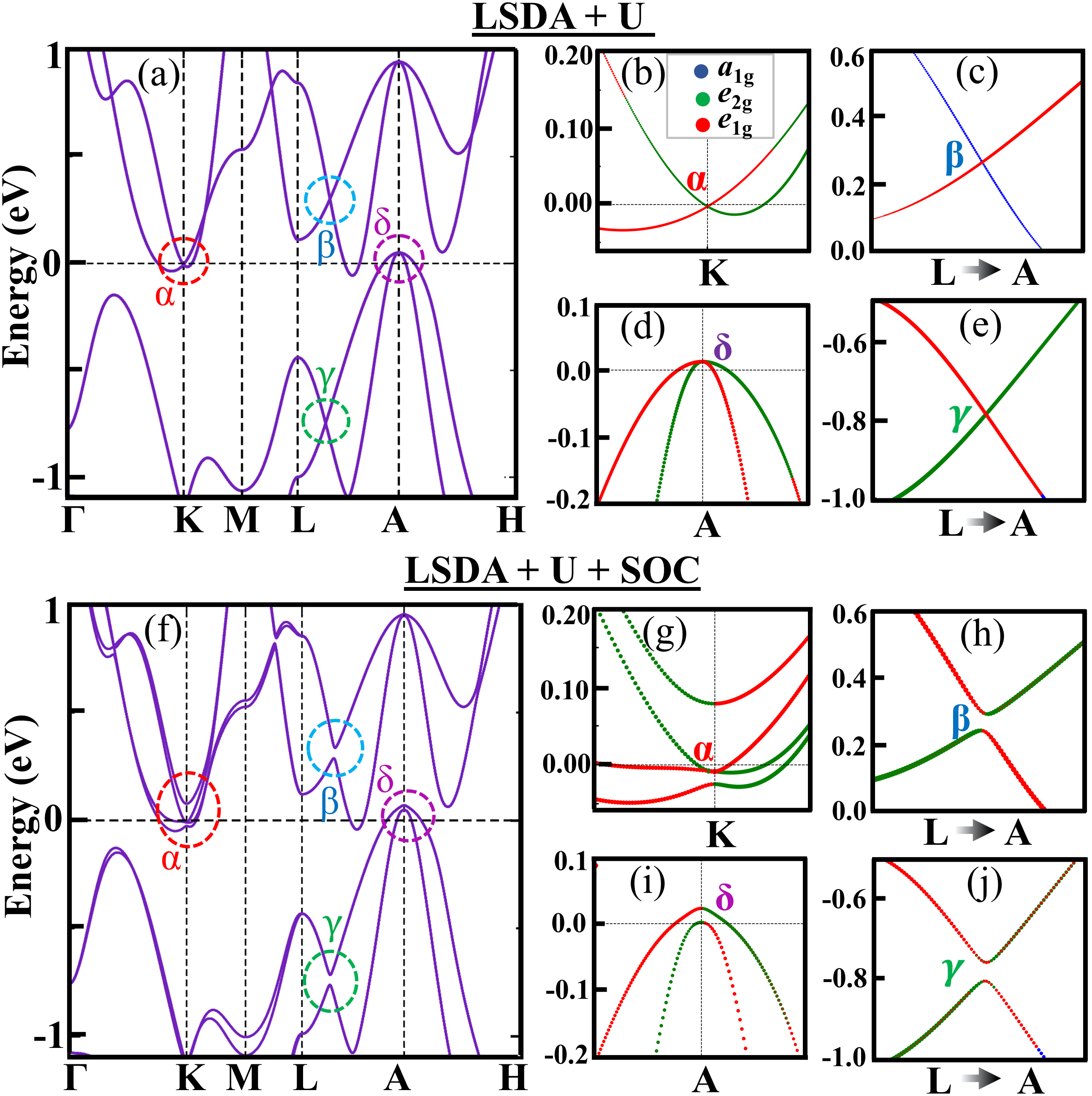}
\caption{(a),(f) Band dispersion along the $\Gamma$–K–M–L–A–H path using LSDA+$U$ and LSDA+$U$+SOC approaches. (b)–(e) Orbital-projected band structures near the Dirac-like crossings ($\alpha$, $\beta$, $\gamma$, $\delta$) along K, A, and L–A. (g)–(j) Corresponding SOC-included orbital projected bands showing lifting of degeneracies. Colored bands denote $a_{1g}$, $e_{2g}$, and $e_{1g}$ characters. Fermi energy ($E_F$) is set at 0 eV.} 
\label{band}
\end{figure}
\par 
\noindent\textbf{Electronic structure:} We first determine the low-temperature magnetic ground state of NiS using DFT within the local spin-density approximation supplemented by a suitable Hubbard $U$ term and SOC (LSDA+U+SOC approach). In agreement with earlier studies~\cite{mypaper}, the magnetic ground state is A-type antiferromagnetic, where ferromagnetically aligned hexagonal Ni layers are coupled antiferromagnetically along the $c$ axis (see Fig. 1(b) of SM~\cite{SM}). The calculated spin and orbital moments on Ni are $\mu_S = 0.90~\mu_B$ and $\mu_L = 0.10~\mu_B$, respectively, yielding $\mu_L/\mu_S \approx 0.11$ and highlighting a non-negligible spin–orbit coupling (SOC) contribution. In this phase, NiS remains semimetallic with a small but finite density of states at the $E_F$, dominated by Ni-$3d$ states (Fig. 2(c) of SM~\cite{SM}). Analysis of the Ni-$3d$ crystal-field splitting using Wannier-based onsite energies and orbital-projected density of states (see Fig. 2 of SI) shows that, due to the hexagonal $P6_3/mmc$ symmetry, the $d$ manifold splits into $a_{1g}$, $e_{2g}$, and $e_{1g}$ which are representations of $D_{3d}$ symmetry. The $a_{1g}$ state lies lowest in energy, followed by $e_{2g}$ and $e_{1g}$, with splittings of $\sim 204$~meV and $\sim61$~meV, respectively (Fig. 2(b) of SM~\cite{SM}). Consequently, the low-energy electronic structure is dominated by the effective $e_{1g}$ and $e_{2g}$ orbitals, which govern the semimetallic and topological properties discussed below. The computed band dispersions using LSDA+$U$ and LSDA+$U$+SOC approaches are shown in Fig.~\ref{band}. A prominent feature of the band structure (Fig.~\ref{band}(a)) is the coexistence of nearly compensated electron and hole pockets with electron pockets near the $K$-point and along the $A$-$L$ line and hole pockets at $A$-point. However, there is another salient but previously unrecognized aspect of the band dispersion. The low-energy spectrum exhibits multiple band crossings, labeled $\alpha$, $\beta$, $\gamma$, and $\delta$ as shown in Fig.~\ref{band}. In the absence of SOC (Figs.~\ref{band}(a–e)), the crossing at the $K$ point ($\alpha$) is fourfold degenerate. This degeneracy originates from two distinct ingredients: (i) spin degeneracy arising from spin-rotational symmetry in LSDA+$U$ approach, and (ii) the two-dimensional irreducible representations of the little group at $K$, inherited from the non-symmorphic space group $P6_3/mmc$ (No.~194). For the A-type antiferromagnetic state with spins aligned along the $c$ axis, time-reversal symmetry $\mathcal{T}$ is broken, whereas inversion symmetry $\mathcal{P}$ remains preserved since the full crystal and magnetic structure are invariant under inversion. In addition, the antiferromagnetic order respects the antiunitary symmetry $\mathcal{S}=\mathcal{T}\tau_{1/2}$, where $\tau_{1/2}$ denotes translation by $c/2$. Consequently, Kramers degeneracy is enforced only on the $k_z=0$ plane, while bands at generic $k_z$ are not symmetry constrained to remain doubly degenerate. Without SOC, the fourfold node at $K$ therefore results from the direct product of spin degeneracy and crystalline symmetry. Similar symmetry-enforced band crossings give rise to the features labeled $\beta$, $\gamma$, and $\delta$ along the $L$–$A$ line and at $A$. Upon inclusion of SOC (Figs.~\ref{band}(f–j)), spin-rotational symmetry is lifted while $\mathcal{P}$ and $\mathcal{S}$ remain intact. The fourfold crossing at $K$ splits into two doubly degenerate bands, reflecting the reduction from spin SU(2) symmetry to the magnetic double group. Crossings not protected by the little-group symmetry acquire small gaps, as visible for $\alpha$, $\beta$, $\gamma$, and $\delta$, whereas double degeneracy survives only where mandated by $\mathcal{S}$ (notably on the $k_z=0$ plane). 
Orbital-resolved analysis indicates that the crossing bands near $E_F$ predominantly derive from Ni $e_{1g}$ and $e_{2g}$ states belonging to distinct irreducible representations. Their inversion near $E_F$ signals a symmetry-distinct band inversion. Since inversion symmetry is preserved, the parity character of the involved states remains well defined, providing a robust symmetry framework for diagnosing possible topological phases. The interplay of non-symmorphic symmetry, preserved $\mathcal{P}$, and antiferromagnetic $\mathcal{S}$ symmetry thus governs the evolution of the Dirac-like crossing at $K$ and its SOC-induced splitting, establishing NiS as a promising antiferromagnetic semimetal platform for topology-driven transport phenomena in correlated $3d$ systems.
\begin{figure}
\centering
\includegraphics[width=1.0\columnwidth]{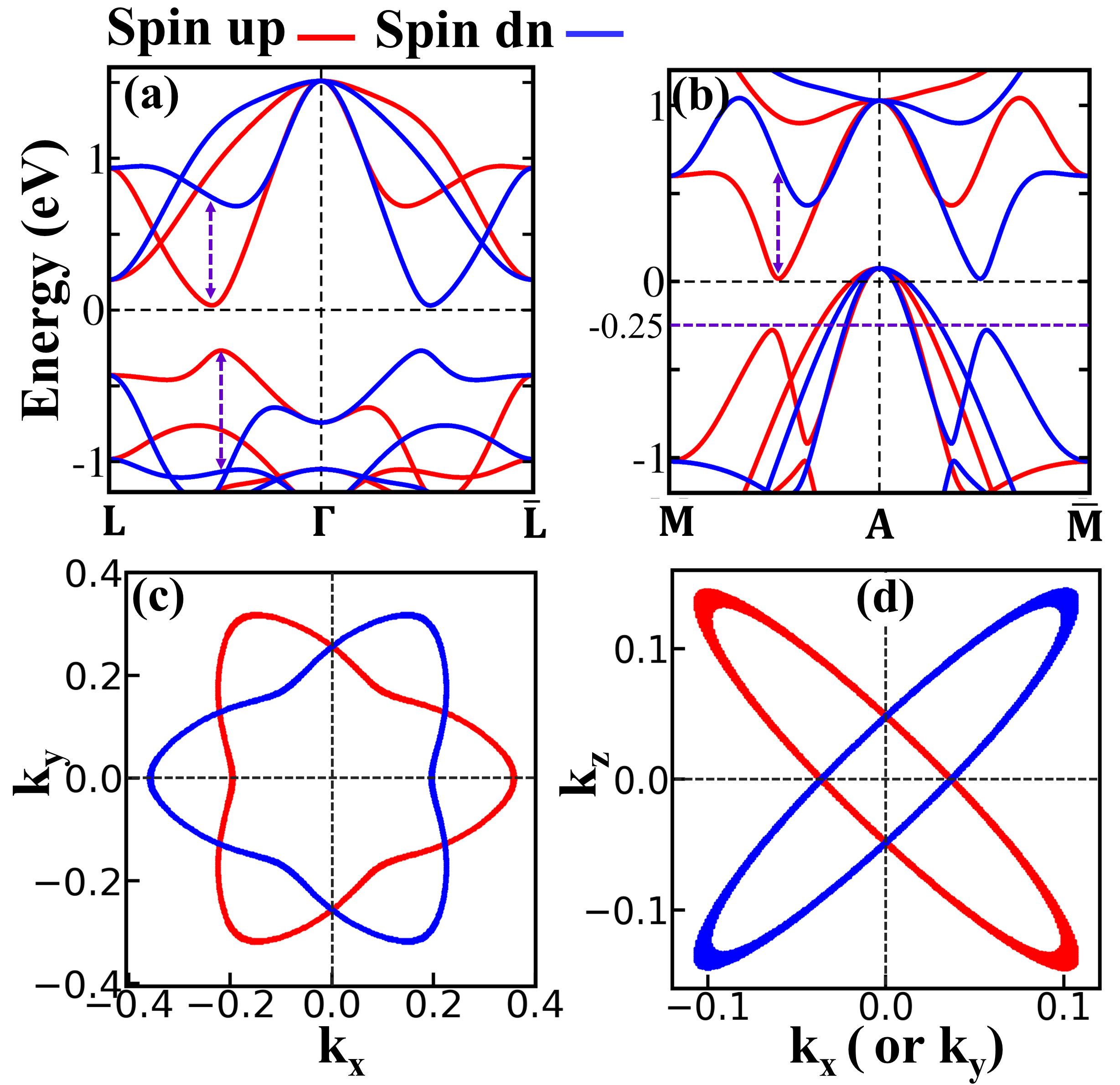}
\caption{Spin-resolved band dispersions along the (a) $L$–$\Gamma$–$\bar{L}$ and (b) $M$–$A$–$\bar{M}$ directions are shown, revealing the characteristic momentum-dependent spin splitting of the altermagnetic state. The Spin-resolved constant-energy contours at $E-E_F=-0.25$ eV in the (c) $k_x$–$k_y$ and (d) $k_x$–$k_z$ (or $k_y$-$k_z$) planes passing through the vicinity of the $A$ point, respectively.} 
\label{bands-alt}
\end{figure}
\par 
Another defining feature of the low-energy electronic structure of NiS is the emergence of momentum ($k$) dependent spin splitting characteristic of altermagnetism, consistent with the symmetry analysis reported in Ref.~\cite{Altermagnet-NiS}. To demonstrate this behavior, we computed the spin-resolved band dispersion along the $L$–$\Gamma$–$\bar{L}$ and $M$–$A$–$\bar{M}$ directions, as shown in Fig.~\ref{bands-alt}(a,b). The corresponding $k$-space paths are highlighted in the Brillouin zone shown in Fig.~1(d) of the SM~\cite{SM}. The pronounced $k$-dependent splitting between the spin-up and spin-down bands emerges away from symmetry-protected momenta, reaching a maximum value of approximately $0.7$ eV along the $L$–$\Gamma$ direction, as indicated by the arrow in Fig.~\ref{bands-alt}(b). This energy scale is comparable to that reported for prototypical altermagnets such as RuO$_2$ and CrSb, where the characteristic spin splitting ranges from $\sim 0.4$ to $1.0$ eV~\cite{RuO2-alt1,RuO2-alt2,CrSb-alt}. Unlike conventional collinear antiferromagnets, where combined $\mathcal{PT}$ or translation-time-reversal symmetries preserve spin degeneracy throughout the Brillouin zone, the A-type AFM order in NiS breaks global time-reversal symmetry while relating the two magnetic sublattices through crystal rotations rather than translations. Consequently, the spin-up and spin-down bands transform into one another under rotational operations of the hexagonal lattice, giving rise to the characteristic anisotropic altermagnetic splitting in momentum space. This effect is particularly pronounced along the $L$–$\Gamma$–$\bar{L}$ and $M$–$A$–$\bar{M}$ directions because these momentum paths are not protected by the antiunitary symmetry $\mathcal{S}=\mathcal{T}\tau_{1/2}$, in contrast to the $k_z=0$ plane where Kramers-like degeneracy remains preserved. 
The altermagnetic nature of the electronic structure is further visualized through the spin-resolved constant-energy contours at $E-E_F=-0.25$ eV (marked by the dashed line in Fig.~\ref{bands-alt}(b)). The iso-energy contours are computed in the $k_x$–$k_y$ and $k_x$–$k_z$ (or $k_y$-$k_z$) planes passing through the vicinity of the $A$ point (see the planes in Fig.~1(d) of the SM~\cite{SM}) and are displayed in Fig.~\ref{bands-alt}(c,d), respectively. In the $k_x$–$k_y$ plane, the spin-up and spin-down states form distinct sixfold-anisotropic contours related by crystal rotation, while in the $k_x$–$k_z$ plane they exhibit elliptical pockets with opposite orientation for opposite spins. These momentum-space textures directly reflect the rotational-coset symmetry of the altermagnetic state and demonstrate the existence of substantial spin splitting despite the absence of net magnetization. The presence of such strong spin-splitting naturally generates strong Berry-curvature hot spots, which form the microscopic origin of the finite anomalous Hall conductivity and large spin Hall response discussed in the following section.
\par
\begin{figure}
\centering
\includegraphics[width=1.0\columnwidth]{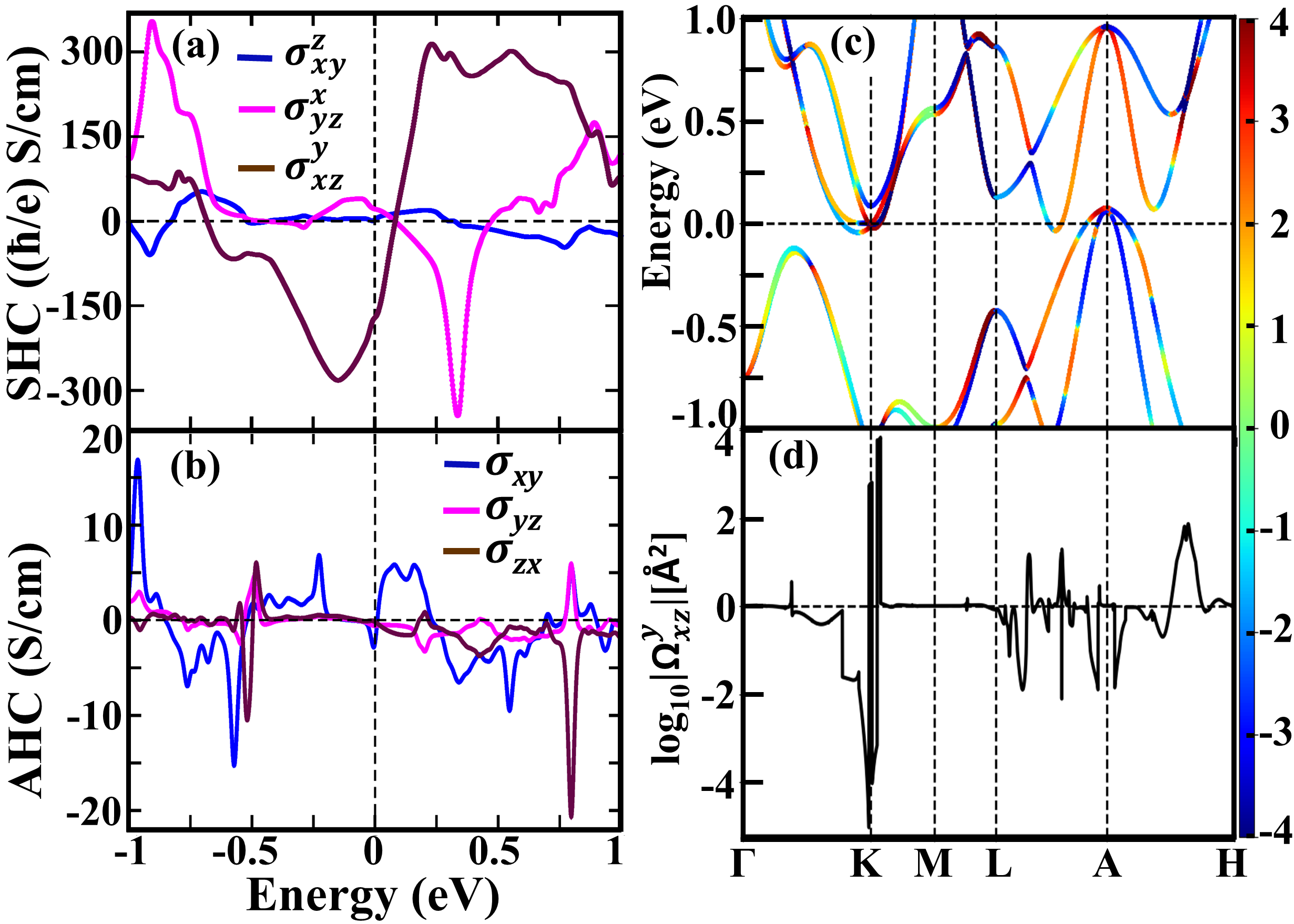}
\caption{Energy (chemical potential) dependence of the (a) SHC and (b) AHC tensor components. (c) Band structure projected with spin Berry curvature (log scale) and (d) $k$-resolved spin Berry curvature at $E=E_F$.} 
\label{shc}
\end{figure}

 
\noindent\textbf{Spin and anomalous Hall conductivity:}
The non-trivial interplay between the preserved inversion symmetry $\mathcal{P}$, broken time-reversal symmetry $\mathcal{T}$, the antiunitary symmetry $\mathcal{S}=\mathcal{T}\tau_{1/2}$ associated with the $c/2$ translation, SOC and momentum-dependent spin splitting discussed above has profound consequences for the spin-dependent transport properties of NiS. For spins aligned along the $c$ axis, the magnetic Laue group remains $6/mmm$, which constrains the spin  tensor $\sigma_{\alpha\beta}^{\gamma}$ to six independent nonzero components. The SHC is third-order tensor whose element $\sigma_{\alpha\beta}^{\gamma}$ quantifies the spin current flowing along the $\alpha$ direction in response to an applied electric field in the $\beta$ direction, with spin polarization along $\gamma$. 
The calculated SHC components are summarized in Fig.~\ref{shc}(a). At $E_F$, the dominant contribution is $\sigma_{xz}^{y} = -171$ $(\hbar/e)$~S/cm, while the remaining tensor elements are smaller in magnitude. The SHC exhibits pronounced energy dependence: $\sigma_{xz}^{y}$ reaches extrema of $-282$ and $314$ $(\hbar/e)$~S/cm at $-0.15$ and $0.23$ eV, respectively, and $\sigma_{yz}^{x}$ attains $-320$ $(\hbar/e)$~S/cm at $0.30$ eV. This strong variation with chemical potential indicates that moderate carrier doping can effectively tune the spin-current response. Importantly, these values exceed those reported for several well-known antiferromagnetic alloys such as PtMn ($\sim 183$ $(\hbar/e)$~S/cm) and IrMn ($\sim 41$ $(\hbar/e)$~S/cm)~\cite{PhysRevLett.113.196602}, and are comparable to peak SHC values found in topological semimetals such as MoTe$_2$ ($\sim -361$ $(\hbar/e)$~S/cm) and WTe$_2$ ($\sim -204$ $(\hbar/e)$~S/cm)~\cite{PhysRevB.99.060408}. These results establish NiS as a $3d$ antiferromagnet hosting an intrinsically large and highly tunable spin Hall response. The microscopic origin of the large SHC is clarified by the momentum-resolved spin Berry curvature for $\sigma_{xz}^y$ as shown in Fig.~\ref{shc} (c) and (d). The color scale (red/blue) represents positive/negative contributions on a logarithmic scale. The dominant contribution originates from SOC-induced avoided crossings near the $K$ point and along the $L$–$A$ line, where Dirac-like band crossings appear in the absence of SOC. These avoided degeneracies act as intense sources of spin Berry curvature. The resulting SHC is therefore a purely relativistic, yet symmetry-enabled, manifestation of the SOC-gapped Dirac crossings discussed above. These results highlight a general principle: in compensated collinear antiferromagnets, symmetry may suppress charge Hall responses while permitting large intrinsic SHC. 
\par 
Remarkably, the same Berry-curvature hot spots that generate the large SHC also produce a sizable intrinsic AHC, as shown in Fig.~\ref{shc}(b). The existence of AHC in NiS could be attributed to its altermagnetic symmetry~\cite{Altermagnet-NiS} as discussed above. As established for the NiAs-type altermagnets~\cite{Altermagnet-NiAs1}, the parent point group is $G=6/mmm$ with halving subgroup $H=\bar{3}m$. The coset $G-H$ is generated by a proper rotational symmetry (sixfold screw $C_{6z}$), not by inversion. Consequently, while inversion symmetry $\mathcal{P}$ is preserved, no global $\mathcal{PT}$ symmetry exists, and time-reversal symmetry is not restored in momentum space. Because AHC transforms as an axial pseudovector, its allowed direction is dictated by the magnetic Laue group. For the A-type order with spins along $c$, the symmetry permits a Hall vector parallel to the $c$ axis. Inversion symmetry enforces $\Omega(\mathbf{k})=\Omega(-\mathbf{k})$, while the antiunitary symmetry $\mathcal{S}=\mathcal{T}\tau_{1/2}$ constrains Berry curvature only on the $k_z=0$ plane. Importantly, no symmetry operation enforces $\Omega(\mathbf{k})=-\Omega(-\mathbf{k})$ over the entire Brillouin zone. Therefore, the Brillouin-zone integral of Berry curvature is not symmetry-forced to vanish, and a finite intrinsic AHC is allowed despite zero net magnetization. The calculated AHC is close to zero at $E_F$, reflecting the compensation of positive and negative Berry-curvature contributions from different FS sheets. However, away from $E_F$, the cancellation is lifted and the AHC reaches values of order $15-20~\mathrm{S/cm}$ around $-0.5$~eV and $+0.8$~eV. The pronounced energy dependence and sign reversal are characteristic of a Berry-curvature–driven intrinsic mechanism. Notably, the magnitude is comparable to that reported for hexagonal noncollinear magnets such as Mn$_3$Ga, where the intrinsic AHC peaks at $\sim 17~\mathrm{S/cm}$~\cite{AHC-Mn3Ga}. This behavior contrasts with collinear antiferromagnets possessing effective $\mathcal{PT}$ or $\mathcal{T}\tau$ symmetry, where Berry curvature cancels identically and intrinsic AHC is strictly forbidden. In NiS, the rotational coset symmetry intrinsic to the altermagnetic NiAs lattice already removes the global cancellation constraint, enabling finite AHC without canting or symmetry reduction. Therefore our analysis provides a correlated $3d$ platform in which compensated collinear order supports both spin and charge Hall responses.  

\begin{figure}
\centering
\includegraphics[width=1\columnwidth]{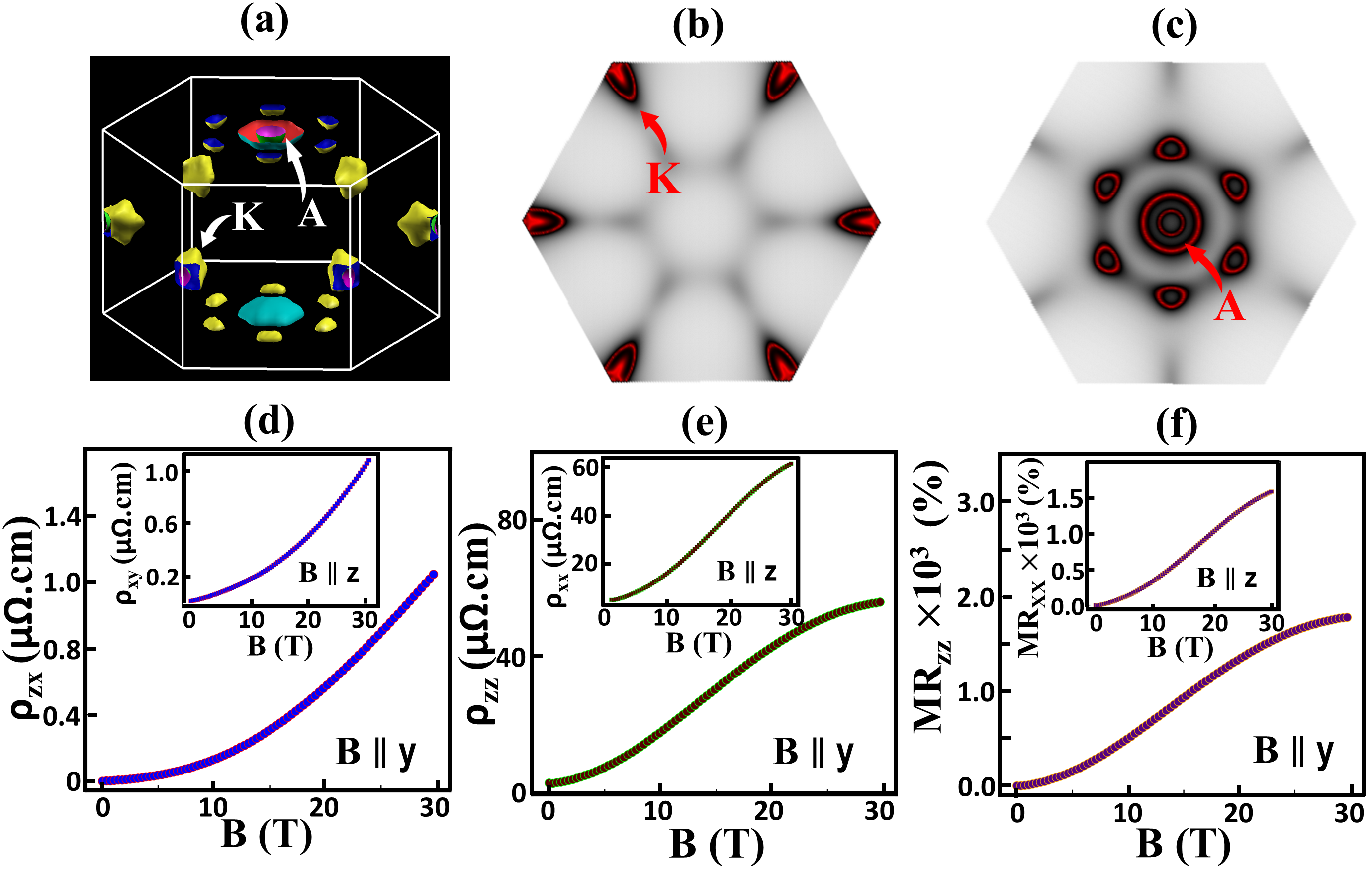}
\caption{(a) Fermi surface (FS) in the hexagonal Brillouin zone, with corresponding cross-sectional cuts in the (b) $k_z=0$ and (c) $k_z=\pi$ planes. Magnetic-field ($B$) dependence of (d) Hall resistivity $\rho_{zx}$, (e) longitudinal resistivity $\rho_{zz}$, and (f) magnetoresistance $MR_{zz}$. Insets in (d)–(f) show $\rho_{xy}$, $\rho_{xx}$, and $MR_{xx}$, respectively.}
\label{MR}
\end{figure}
\par 
\noindent\textbf{Fermi-surface topology and large magnetoresistance:}
Beyond the Berry-curvature–driven Hall responses, the electronic structure of NiS leaves a distinct imprint on its magnetotransport properties. As evident from the band structure, multiple bands cross the $E_F$, giving rise to both electron- and hole-like carriers. The corresponding three-dimensional FS is shown in Fig.~\ref{MR}(a), with two-dimensional projections in the $k_z=0$ and $k_z=\pi$ planes displayed in Figs.~\ref{MR}(b) and \ref{MR}(c). Electron pockets are centered near the $K$ points, while a hole pocket forms around $A$. The comparable pocket volumes yield nearly equal carrier densities ($n_e \approx n_h \approx 2.73\times10^{20}$~cm$^{-3}$), indicating strong electron–hole compensation. Such electron–hole compensation is a well-established microscopic origin of the large, nonsaturating magnetoresistance observed in materials such as WTe$_2$~\cite{Ali_WTe2}, NbP~\cite{Xu_NbP}, LaBi~\cite{Tafti_LaBi}, and ZrSiS~\cite{Schoop_ZrSiS}. In such cases, opposite Lorentz forces acting on electrons and holes suppress transverse conductivity under magnetic field, producing a large MR that persists to high fields. To quantify the field response in NiS, we computed the conductivity tensor within semiclassical Boltzmann transport including SOC. The transverse resistivity $\rho_{zx}$ (Fig.~\ref{MR}(d)) exhibits pronounced nonlinear field dependence, characteristic of multicarrier compensated transport. The longitudinal resistivity $\rho_{zz}$ (Fig.~\ref{MR}(e)) increases monotonically from $\sim 3~\mu\Omega$\,cm at zero field to $\sim 60~\mu\Omega$\,cm at $30$~T (50~K). As a result, the magneto-resistance ($MR_{zz}$ in Fig.~\ref{MR}(f)) reaches $\sim 1.7\times10^3$\% at $30$~T, in good agreement with experimental reports of MR $\sim 1500$\% in bulk NiS at lower fields~\cite{mr2}.  Similar behavior is also observed in $\rho_{xx}$ and $MR_{xx}$ as shown in the inset of Fig.~\ref{MR}(e) and (f). The large magnitude and nonsaturating character of the MR place NiS among compensated semimetals with strong magnetotransport responses. What makes NiS particularly notable, however, is that this large compensation-driven MR coexists with sizable intrinsic SHC and symmetry-allowed AHC originating from altermagnetic spin splitting and SOC-gapped Dirac-like crossings. In most systems, large MR occurs in nonmagnetic semimetals, while strong Berry-curvature–driven Hall effects typically appear in magnetic materials without near-perfect compensation. The simultaneous realization of (i) compensated electron–hole transport, (ii) large intrinsic SHC, and (iii) finite intrinsic AHC in a chemically simple $3d$ altermagnet is therefore highly uncommon. NiS thus exemplifies a multifunctional transport platform in which semiclassical carrier compensation and quantum Berry curvature effects coexist within the same band structure. This combination is of considerable interest for spintronic and magnetotransport applications, where large field sensitivity, intrinsic spin-current generation, and magnetic symmetry control can be harnessed within a single correlated $3d$ compound. 
\par
\begin{figure}
\includegraphics[width=1\columnwidth]{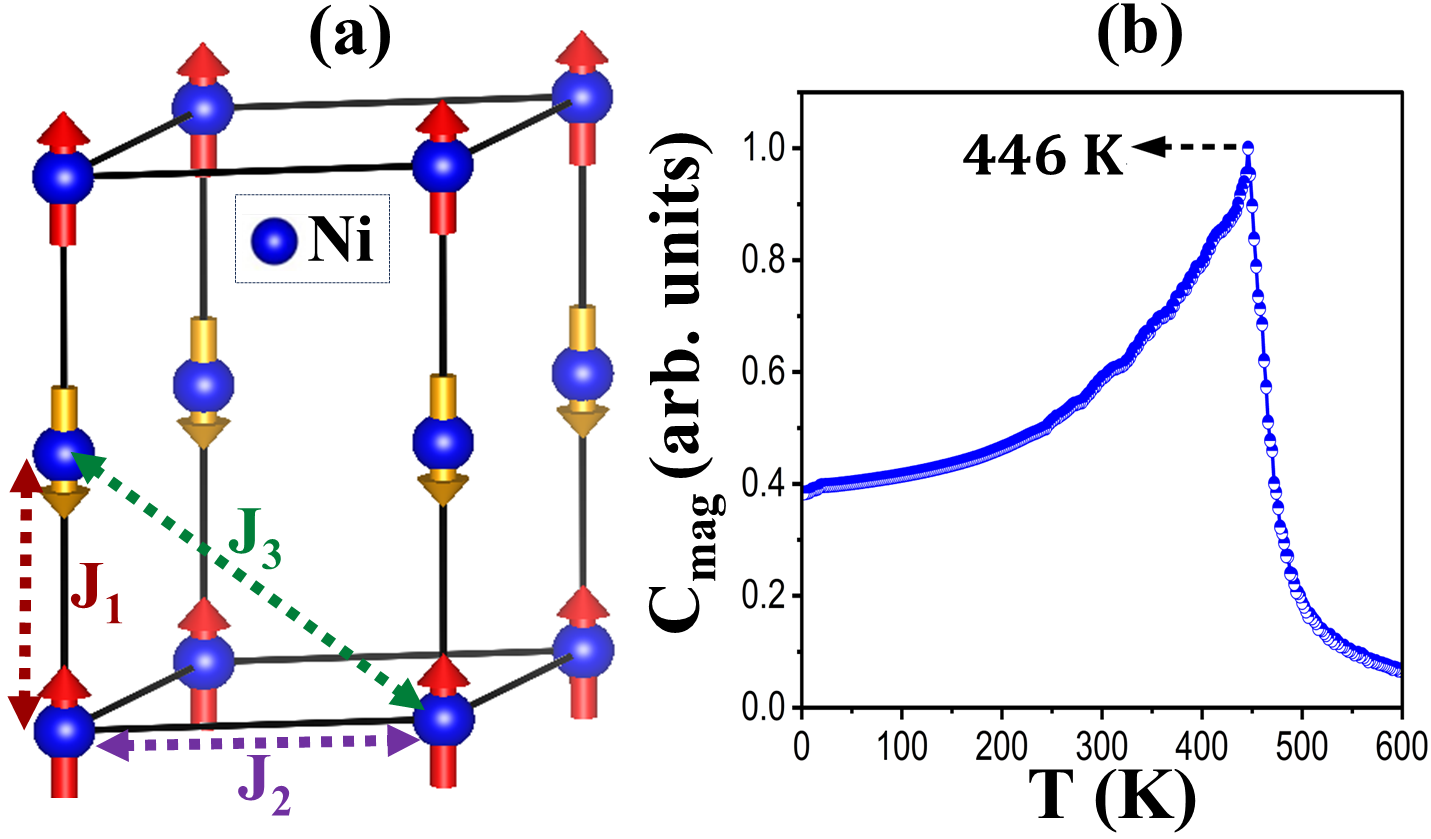}
\caption{(a) A-type antiferromagnetic structure of NiS illustrating the dominant exchange interactions $J_1$, $J_2$, and $J_3$ (only Ni ions are shown for clarity). (b) Calculated magnetic specific heat as a function of temperature, displaying a pronounced anomaly at the N\'eel temperature $T_N \approx 446$ K.} 
\label{Sp-ph}
\end{figure}

 
\noindent\textbf{Spin Hamiltonian:}
The electronic and transport properties discussed above including the altermagnetic spin splitting, Dirac-like band crossings, Berry-curvature-driven Hall responses, and compensated semimetallic behavior, are realized within the A-type AFM ground state of NiS. A microscopic understanding of the magnetic interactions is therefore essential, both to establish the stability of this magnetic phase and to identify the exchange mechanisms that govern the underlying altermagnetic electronic structure. Hence,  we computed the magnetic exchange interactions and SOC-induced anisotropic couplings using the Korringa–Kohn–Rostoker Green-function formalism implemented in the RSPt code~\cite{magneticforceth1,RevModPhys.95.035004,PhysRevB.103.174422,magneticforceth2}. Three dominant isotropic interactions emerge: the nearest-neighbor ($J_1$), second-neighbor ($J_2$), and third-neighbor ($J_3$) couplings, as illustrated in Fig.~\ref{Sp-ph}(a). As shown in Table~\ref{tab:mag_params}, both $J_1$ and $J_3$ are strongly antiferromagnetic, while $J_2$ is weakly ferromagnetic. Notably, the longer-range $J_3$ exceeds $J_1$ in magnitude despite the larger Ni–Ni distances (see SM~\cite{SM}), indicating an efficient superexchange pathway. This hierarchy follows directly from the Ni–S–Ni bonding geometry and the $d^8$ configuration of Ni$^{2+}$. According to the Goodenough–Kanamori–Anderson rules, the near-$90^\circ$ bond angle for the second neighbor ($\sim91^\circ$) favors weak ferromagnetic exchange, whereas the larger deviations from $90^\circ$ for the first ($68^\circ$) and third ($131^\circ$) neighbors enhance antiferromagnetic $e_g$–$e_g$ superexchange. The SOC introduces additional anisotropic terms. While inversion symmetry suppresses DM interactions for the first and second NN, the third NN path lacks an inversion center and supports a finite DM vector with a magnitude of $D_3 =0.25 $~meV. In addition, the single-ion anisotropy ($\varepsilon_i^{\rm AN}$ = 0.14 meV/Ni) is sizable, reflecting a significant SOC-induced magnetocrystalline anisotropy that stabilizes the crystallographic $c$ axis as the easy axis of magnetization.
Collectively, these terms define the following effective spin Hamiltonian,
\begin{equation}
H_{\mathrm{spin}} = -\frac{1}{2}\sum_{i\neq j} 
J_{ij}\,\vec{S}_i\!\cdot\!\vec{S}_j
-\sum_{i\neq j}\mathbf{D}_{ij}\!\cdot\![\vec{S}_i\times\vec{S}_j]
-\sum_i \varepsilon_i^{\rm AN} (S_i^z)^2 .
\end{equation}
The sLLG simulations based on this $H_{spin}$ reproduce a sharp antiferromagnetic transition at $T_N \approx 446$~K as shown in Fig.~\ref{Sp-ph}(b). This is in excellent agreement with experiment~\cite{NiS-Tn1}, confirming that the first-principles parameters faithfully describe the magnetism of NiS on which the electronic topology and transport phenomena discussed above are based.
\begin{table}[t]
\centering
\caption{Isotropic exchange interactions ($J_1$, $J_2$, $J_3$), components of the DM vector for third NN ($\mathbf{D}_3$), and  single-ion anisotropy ($\varepsilon_i^{\rm AN}$) per Ni. All values are in meV.}
\renewcommand{\arraystretch}{1.3}
\setlength{\tabcolsep}{7pt}
\begin{tabular}{lcccccccc}
\hline \hline
 & $J_1$ & $J_2$ & $J_3$ & 
 \multicolumn{3}{c}{$\mathbf{D}_3$} & $\varepsilon_i^{\rm AN}$ \\
 &  &  &  & $D_x$ & $D_y$ & $D_z$ &  \\
\midrule
LT & -3.31 & 0.69 & -4.53 & 0.14 & 0.14 & 0.15 & 0.14 \\
\bottomrule
\end{tabular}
\label{tab:mag_params}
\end{table}
\par 
\noindent\textbf{Conclusion:}
In summary, we establish NiS as a correlated $3d$ altermagnetic semimetal where Dirac-like topology, carrier compensation, Berry-curvature-driven transport, and magnetic order emerge from a common symmetry-governed electronic structure. The rotational coset symmetry of the NiAs lattice produces momentum-dependent spin splitting characteristic of altermagnetism. Upon inclusion of SOC, the resulting gapped Dirac-like crossings generate pronounced Berry-curvature hot spots, yielding large intrinsic SHC and symmetry-allowed AHC despite zero net magnetization. Simultaneously, nearly perfect electron–hole compensation gives rise to nonsaturating magnetoresistance of order $10^3$\%, placing NiS among compensated semimetals with strong field response. Unlike rare-earth monopnictides and nonmagnetic semimetals such as LuBi, YBi, NdBi, PrBi, and ScSb~\cite{LuBi-YBi,NdBi,PrBi,ScSb1,ScSb2} where extreme magnetoresistance typically arises from semiclassical compensation alone or heavy-element systems such as $\alpha$-WP$_2$ and NbSb$_2$~\cite{WP2,NbSb2} in which strong SOC dominates transport, NiS realizes a distinct regime where compensation-driven magnetoresistance and Berry-curvature–mediated Hall effects coexist in a chemically simple $3d$ altermagnet. This combination is exceptionally rare and points to a more general design principle in which compensated FS, symmetry-enforced band crossings, and superexchange collectively enable multifunctional behavior in correlated $3d$ systems. On the magnetic side, first-principles derived spin Hamiltonian quantitatively reproducing the experimental N\'eel temperature. Thus, our work establishes transition-metal chalcogenides as a promising platform for multifunctional quantum responses, demonstrating the interplay of Dirac-like topology, altermagnetism, and compensated transport, opening new avenues for symmetry-controlled quantum functionalities.
\section{ACKNOWLEDGMENTS}
S.K.P. acknowledges support from ANRF (previously
SERB), Government of India for the core research grant
(CRG/2023/003063).
\section{DATA AVAILABILITY}
The data that support the findings of this article are available
from the authors upon reasonable request.

\clearpage
\appendix

\section*{Supplementary Information}

\setcounter{section}{0}
\renewcommand{\thesection}{S\arabic{section}}
\renewcommand{\thefigure}{S\arabic{figure}}
\renewcommand{\thetable}{S\arabic{table}}
\renewcommand{\theequation}{S\arabic{equation}}

\setcounter{figure}{0}
\setcounter{table}{0}
\setcounter{equation}{0}

\subsection*{LSDA+$U$+ SOC calculations}
First-principles electronic-structure calculations were carried out within density functional theory (DFT)~\cite{DFT1,DFT2} using the full-potential linearized augmented plane-wave (FP-LAPW) method, as implemented in the WIEN2k code~\cite{wien2k}. The muffin-tin radii were chosen as 2.32~a.u.\ for Ni and 2.06~a.u.\ for S. Wave functions were expanded in plane waves in the interstitial region and atomic-like basis functions inside the muffin-tin spheres, with additional local orbitals for Ni-$3p$ and S-$3s$ states included to improve the description of semicore levels. The plane-wave cutoff was set by $R_{MT}k_{\mathrm{max}}=7$, where $R_{MT}$ is the smallest muffin-tin radius and $k_{\mathrm{max}}$ is the maximum reciprocal-lattice vector. The charge density was expanded up to $G_{\mathrm{max}}=12$. Brillouin-zone integrations were performed using the tetrahedron method~\cite{tetrahedron} with a dense $k$-mesh corresponding to 885 $k$ points in the irreducible Brillouin zone, ensuring well-converged total energies. Exchange-correlation effects were treated within the local spin-density approximation (LSDA), augmented by an on-site Hubbard interaction ($U$) for the Ni-$3d$ states within the LSDA+$U$ formalism to account for electronic correlations. The value of $U=2.3$~eV was adopted based on earlier studies of NiS~\cite{mypaper}. Spin-orbit coupling (SOC) was included self-consistently using the second-variational scheme, which is essential for capturing the spin-dependent band topology and Berry-curvature-related transport responses discussed in this work.
 
\begin{figure*}[t]
\includegraphics[width=2\columnwidth]{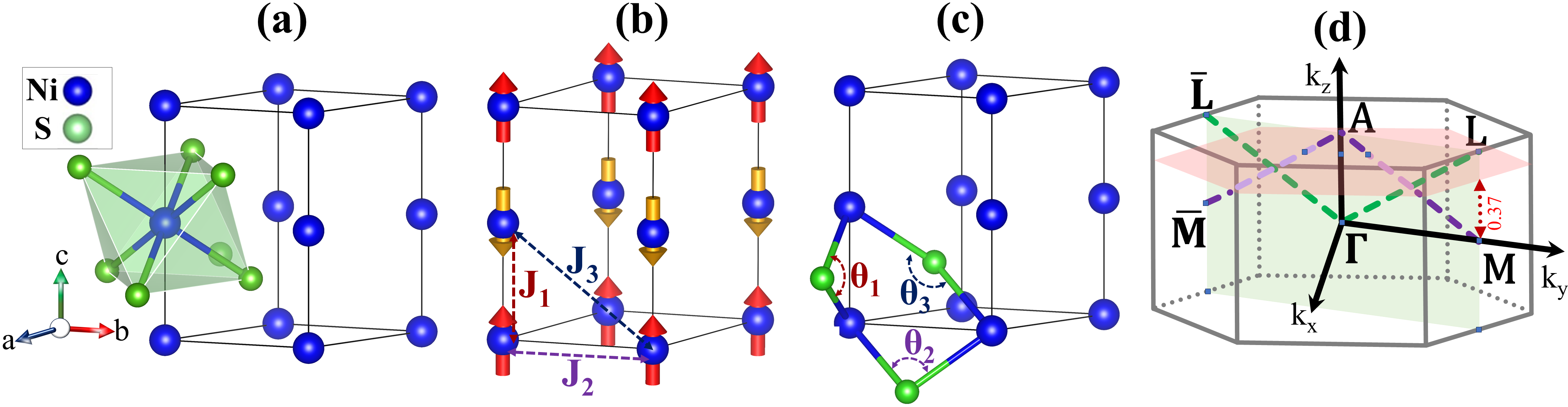}
\caption{(a) Crystal structure of NiS showing the NiS$_6$ octahedral coordination, (b) the symmetry-distinct Ni–Ni exchange pathways $J_{1}$, $J_{2}$, and $J_{3}$ are illustrates with only Ni atoms displayed for clarity. Panel (c) shows the Ni–S–Ni bond angles associated with each exchange path, highlighting the distinct superexchange geometries that govern the magnitude and sign of $J_{ij}$.(d) The altermagnetic $k$-paths $L$-$\Gamma$-$\bar{L}$ and $M$–$A$–$\bar{M}$ are shown with the dotted green and volet line, respectively, within the Hexagonal brillouin zone.
} 
\label{SI-fig1}
\end{figure*}
\subsection*{Berry curvature and intrinsic spin Hall conductivity}
To compute the intrinsic spin Hall effect (SHE), we constructed a tight-binding Hamiltonian based on maximally localized Wannier functions using the Wannier90 package~\cite{w90}. The resulting tight-binding model was then employed within the WannierTools framework~\cite{wanniertool}, where the spin Hall conductivity (SHC) was evaluated using the Green-function implementation of linear-response theory~\cite{wanniertool-green}. The intrinsic SHC was calculated using the Kubo–Berry formalism, which captures the dissipationless contribution originating from the momentum-space spin Berry curvature of the occupied Bloch states~\cite{shc1}. Within this framework, the SHC tensor component $\sigma_{\alpha\beta}^{\gamma}$ is obtained by integrating the spin Berry curvature over the Brillouin zone, ensuring an accurate description of the SOC-induced interband coherence effects responsible for the intrinsic SHE. This approach allows us to directly link the calculated SHC to the SOC-gapped Dirac-like crossings and Berry-curvature hot spots identified in the electronic structure analysis. For a given spin polarization along the $\gamma$ direction, the SHC tensor element $\sigma_{\alpha\beta}^{\gamma}$ is given by
\begin{equation}
\sigma_{\alpha\beta}^{\gamma}
=
-\frac{e^{2}}{\hbar}\,
\frac{1}{V N_k}
\sum_{\mathbf{k}}
\Omega_{\alpha\beta}^{\gamma}(\mathbf{k}),
\label{sigma}
\end{equation}
where $V$ is the volume of the primitive unit cell, $N_k$ is the total number of $\mathbf{k}$ points sampling the Brillouin zone, and $\Omega_{\alpha\beta}^{\gamma}(\mathbf{k})$ denotes the momentum-resolved spin Berry curvature. The total spin Berry curvature at a given $\mathbf{k}$ point is obtained by summing over all occupied bands,
\begin{equation}
\Omega_{\alpha\beta}^{\gamma}(\mathbf{k})
=
\sum_{n}
f_{n\mathbf{k}}\,
\Omega_{n,\alpha\beta}^{\gamma}(\mathbf{k}),
\label{sp-B}
\end{equation}
where $f_{n\mathbf{k}}$ is the Fermi-Dirac distribution function and $n$ labels the band index.

The band-resolved spin Berry curvature is expressed as
\begin{equation}
\Omega_{n,\alpha\beta}^{\gamma}(\mathbf{k})
=
\hbar^{2}
\sum_{m\neq n}
\frac{
-2\,\mathrm{Im}
\left[
\langle n\mathbf{k} |
\hat{J}_{\alpha}^{\gamma}
| m\mathbf{k} \rangle
\langle m\mathbf{k} |
\hat{v}_{\beta}
| n\mathbf{k} \rangle
\right]
}{
\left(\epsilon_{n\mathbf{k}}-\epsilon_{m\mathbf{k}}\right)^{2}
},
\label{eq:spin_berry}
\end{equation}
where $\epsilon_{n\mathbf{k}}$ and $\epsilon_{m\mathbf{k}}$ are the eigenvalues of the Bloch states $|n\mathbf{k}\rangle$ and $|m\mathbf{k}\rangle$, respectively. The spin-current operator is defined as
\begin{equation}
\hat{J}_{\alpha}^{\gamma}
=
\frac{1}{2}
\left\{
\hat{\sigma}_{\gamma},
\hat{v}_{\alpha}
\right\},
\end{equation}
with $\hat{\sigma}_{\gamma}$ the Pauli spin operator and $\hat{v}_{\alpha}$ the velocity operator along the $\alpha$ direction, given by
\begin{equation}
\hat{v}_{\alpha}
=
\frac{1}{\hbar}
\frac{\partial \hat{H}(\mathbf{k})}{\partial k_{\alpha}}.
\label{velocity}
\end{equation}

This formulation explicitly highlights that the intrinsic SHC is governed by interband matrix elements weighted by the inverse square of the energy separation between bands. Consequently, spin-orbit-induced avoided crossings and symmetry-protected near-degeneracies close to the Fermi level generate intense spin Berry curvature. In NiS, these Berry-curvature hot spots are strongly localized near Dirac-like band crossings and nonsymmorphic symmetry-enforced degeneracies, leading to the large and anisotropic SHC reported in the main text.
For numerical evaluation, dense $\mathbf{k}$-point meshes ($251\times251\times251$) were employed to ensure convergence of the Berry curvature, and the summation was restricted to occupied states. The same formalism, with the spin-current operator replaced by the charge-current operator, was used to compute the intrinsic anomalous Hall conductivity. 

\subsection*{Boltzmann transport theory}
To evaluate the magnetotransport properties, we employed the semiclassical Boltzmann transport formalism as implemented in the WannierTools package~\cite{wanniertool}. The calculation is carried out under two standard assumptions: (i) the rigid-band approximation, in which the applied magnetic field does not modify the electronic structure or Fermi surface topology, and (ii) the relaxation-time approximation, where the carrier lifetime $\tau_n$ is taken to be band dependent but independent of crystal momentum $\mathbf{k}$. Within this semiclassical framework, charge carriers follow magnetic-field–driven trajectories in momentum space. Under these assumptions, the conductivity tensor in the presence of a magnetic field $\mathbf{B}$ is expressed as
\begin{equation}
\frac{\sigma_{ij}^{(n)}(B)}{\tau_n}
=
\frac{e^{2}}{4\pi^{3}}
\int d\mathbf{k}\,
v_i^{n}(\mathbf{k})\,
\bar{v}_j^{\,n}(\mathbf{k})
\left(
-\frac{\partial f}{\partial \varepsilon}
\right)_{\varepsilon=\varepsilon_n(\mathbf{k})},
\end{equation}
where $e$ is the electron charge, $\varepsilon_n(\mathbf{k})$ is the band dispersion, $f$ is the Fermi–Dirac distribution function, and the band velocity is given by $\mathbf{v}_n(\mathbf{k})=\nabla_{\mathbf{k}}\varepsilon_n(\mathbf{k})/\hbar$. The quantity $\bar{\mathbf{v}}_n(\mathbf{k})$ denotes the velocity averaged along the semiclassical trajectory $\mathbf{k}_n(t)$ of a carrier in band $n$ under the magnetic field:
\begin{equation}
\bar{\mathbf{v}}_{n}(\mathbf{k})
=
\int_{-\infty}^{0}
\frac{d(Bt)}{B\tau_n}
\, e^{t/\tau_n}
\, \mathbf{v}_n\!\left(\mathbf{k}_n(t)\right).
\end{equation}
This time-averaged velocity accounts for cyclotron motion in the presence of the magnetic field, with scattering effects incorporated through the exponential damping factor governed by $\tau_n$. The total conductivity tensor is obtained by summing over all relevant bands. The resistivity tensor is then computed as the inverse of the conductivity tensor, $\boldsymbol{\rho}=\boldsymbol{\sigma}^{-1}$. The transverse magnetoresistance (MR) is defined as
\begin{equation}
\mathrm{MR}(\%)
=
\frac{\rho(B)-\rho(0)}{\rho(0)} \times 100\%,
\label{MR}
\end{equation}
where $\rho(B)$ and $\rho(0)$ denote the longitudinal resistivity in the presence and absence of magnetic field, respectively.

\begin{figure*}[t]
\includegraphics[width=2\columnwidth]{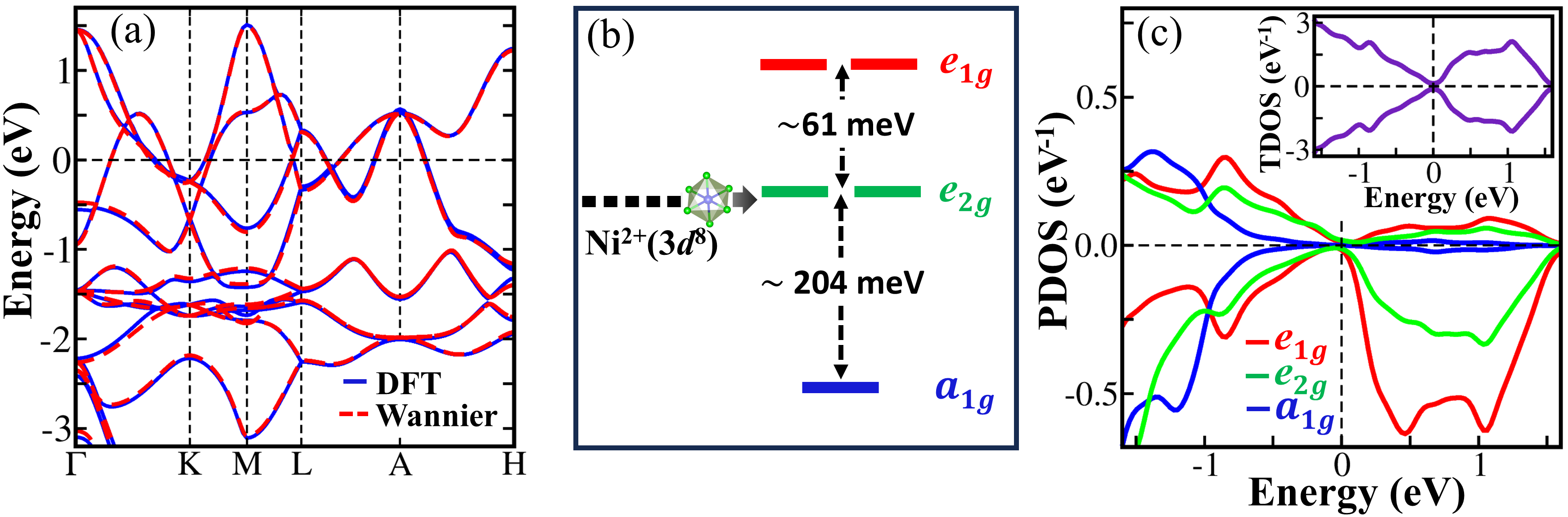}
\caption{(a) Comparison between the non-spin-polarized DFT band structure (blue solid lines) and the Wannier-interpolated bands (red dashed lines) in the vicinity of the Fermi level, demonstrating the accuracy of the Wannier tight-binding model. (b) Schematic illustration of the crystal-field splitting of Ni$^{2+}$ ($3d^8$) states into the symmetry-adapted $e_{1g}$, $e_{2g}$, and $a_{1g}$ manifolds. (c) Orbital-resolved partial density of states (PDOS) for the $e_{1g}$, $e_{2g}$, and $a_{1g}$ states near the Fermi energy obtained from LSDA+$U$+SOC calculations; the corresponding total density of states (TDOS) is shown in the inset.
}
\label{SI-fig2}
\end{figure*}

\subsection*{Magnetic exchange interactions and spin Hamiltonian}
The magnetic exchange interactions were computed using the full-potential linearized muffin-tin orbital (FPLMTO) method as implemented in the \textsc{RSPt} code~\cite{RSPT1,RSPT2,RSPT3}. The isotropic exchange parameters were evaluated within the Lichtenstein–Katsnelson–Antropov–Gubanov (LKAG) formalism based on the magnetic force theorem and Green’s-function approach~\cite{magneticforceth1,magneticforceth2}. In this framework, the intersite exchange coupling between sites $i$ and $j$ is given by
\begin{equation}
J_{ij}=\frac{T}{4}\sum_{n}
\mathrm{Tr}\!\left[
\hat{\Delta}_i(i\omega_n)
\hat{G}_{ij}^{\uparrow}(i\omega_n)
\hat{\Delta}_j(i\omega_n)
\hat{G}_{ji}^{\downarrow}(i\omega_n)
\right],
\end{equation}
where $\hat{\Delta}_i$ is the on-site exchange splitting, $\hat{G}_{ij}^{\sigma}$ denotes the spin-resolved intersite Green’s function, $T$ is the electronic temperature, and $\omega_n$ are fermionic Matsubara frequencies.
\par 
To capture relativistic effects and anisotropic magnetic interactions, such as Dzyaloshinskii–Moriya (DM), we employed the fully relativistic extension of this formalism within the Korringa–Kohn–Rostoker (KKR) Green-function framework~\cite{RevModPhys.95.035004,PhysRevB.103.174422}. In this approach, the magnetic interactions are mapped onto a generalized Heisenberg Hamiltonian,
\begin{equation}
H=-\frac{1}{2}\sum_{i\neq j} J_{ij}^{\alpha\beta} m_i^\alpha m_j^\beta,
\quad \alpha,\beta=x,y,z,
\label{H-exc}
\end{equation}
where $J_{ij}^{\alpha\beta}$ is the full exchange tensor in spin space. In Cartesian coordinates,
\begin{equation}
J_{ij}^{\alpha\beta} =
\begin{pmatrix}
J_{xx} & J_{xy} & J_{xz}\\
J_{yx} & J_{yy} & J_{yz}\\
J_{zx} & J_{zy} & J_{zz}
\end{pmatrix}.
\end{equation}
The isotropic exchange interaction is obtained from the trace,
\begin{equation}
J_{ij}=\frac{1}{3}(J_{xx}+J_{yy}+J_{zz}),
\end{equation}
while the DM vector $\mathbf{D}_{ij}$ is extracted from the antisymmetric part of the tensor,
\begin{equation}
\begin{aligned}
D_{ij}^{x} &= \frac{1}{2}\left( J_{ij}^{yz} - J_{ij}^{zy} \right), \\
D_{ij}^{y} &= \frac{1}{2}\left( J_{ij}^{zx} - J_{ij}^{xz} \right), \\
D_{ij}^{z} &= \frac{1}{2}\left( J_{ij}^{xy} - J_{ij}^{yx} \right).
\end{aligned}
\end{equation}

Using these parameters, the magnetic interactions are described by the effective spin Hamiltonian
\begin{equation}
H_{\mathrm{spin}} = -\frac{1}{2}\sum_{i\neq j} 
J_{ij}\,\vec{S}_i\!\cdot\!\vec{S}_j
-\sum_{i\neq j}\mathbf{D}_{ij}\!\cdot\![\vec{S}_i\times\vec{S}_j]
-\sum_i \varepsilon_i^{\rm AN} (S_i^z)^2 ,
\label{Hspin}
\end{equation}
where $\varepsilon_i^{\rm AN}$ denotes the single-ion anisotropy. This Green-function–based approach has been extensively validated for transition-metal compounds~\cite{LaCrS3,PhysRevB.103.174422,Ir-Ru} and provides a reliable microscopic description of both isotropic and anisotropic magnetic interactions in NiS.
\subsection{Spin-dynamics simulations}
Finite-temperature spin dynamics were simulated by solving the stochastic Landau–Lifshitz–Gilbert (sLLG) equation using the effective spin Hamiltonian [Eq.~(\ref{Hspin})], as implemented in the \textsc{UppASD} code~\cite{UPPASD1,SLLG2}. Simulations were performed on a $20\times20\times20$ supercell with periodic boundary conditions over a temperature range of 0–600~K, covering the experimentally relevant regime. The time evolution of the local magnetic moments was governed by the stochastic Landau–Lifshitz–Gilbert (sLLG) equation~\cite{SLLG1,SLLG2},
\begin{equation}
\frac{\partial \mathbf{m}_i}{\partial t}
=
-\frac{\gamma}{1+\alpha^2}
\left[
\mathbf{m}_i \times \mathbf{B}_i^{\mathrm{eff}}
+
\frac{\alpha}{m}
\mathbf{m}_i \times
\left(
\mathbf{m}_i \times \mathbf{B}_i^{\mathrm{eff}}
\right)
\right],
\end{equation}
where $\gamma$ is the gyromagnetic ratio and $\alpha$ the Gilbert damping parameter. The effective field acting on site $i$ is given by
$\mathbf{B}_i^{\mathrm{eff}}=\mathbf{B}_i+\mathbf{b}_i(t)$.
Here, $\mathbf{B}_i=-\partial H_{\mathrm{spin}}/\partial \mathbf{m}_i$ is the deterministic field derived from the spin Hamiltonian, while $\mathbf{b}_i(t)$ represents a stochastic magnetic field describing thermal fluctuations. The stochastic term is treated as Gaussian white noise and satisfies the fluctuation–dissipation theorem, ensuring proper thermal equilibrium.

The sLLG equation was integrated using the midpoint scheme~\cite{Mentink_2010} with a time step $\Delta t=0.1$~fs. To accelerate thermalization, the damping parameter was set to $\alpha=1.0$ during equilibration. After reaching steady state at each temperature, thermodynamic quantities were extracted from time-averaged observables. The magnetic contribution to the specific heat was computed from energy fluctuations according to
\begin{equation}
C_M =
\frac{\langle E^2 \rangle - \langle E \rangle^2}{k_B T^2},
\end{equation}
where $k_B$ is the Boltzmann constant. The temperature dependence of $C_M(T)$ was analyzed to identify signatures of magnetic phase transitions in NiS.
\section*{Crystal structure and symmetry analysis}
Nickel sulfide (NiS) crystallizes in the hexagonal NiAs-type structure with space group $P6_3/mmc$ (No.~194)~\cite{cryststruct}, as shown in Fig.~\ref{SI-fig1}(a). The corresponding crystallographic point group is $D_{6h}$ ($6/mmm$), which contains inversion symmetry ($\mathcal{P}$), horizontal and vertical mirror planes, twofold rotations, and a non-symmorphic sixfold screw axis $\{C_{6z}|0,0,\tfrac{1}{2}\}$ along the $c$ direction. At high temperature, NiS is in a paramagnetic (PM) phase that preserves the full symmetry of the $P6_3/mmc$ space group together with time-reversal symmetry $\mathcal{T}$. Upon cooling below the magnetic ordering temperature, the system develops long-range magnetic order. In the low-temperature phase, NiS adopts an A-type antiferromagnetic (AFM) configuration (Fig.~\ref{SI-fig1}(b)), which is the focus of the present work.
\par 
In the $P6_3/mmc$ structure, Ni atoms occupy the $2a$ Wyckoff positions $(0,0,0)$ and $(0,0,\tfrac{1}{2})$, forming a hexagonal close-packed (hcp) sublattice. S atoms reside at the $2c$ positions $(\tfrac{1}{3},\tfrac{2}{3},\tfrac{1}{4})$ and $(\tfrac{2}{3},\tfrac{1}{3},\tfrac{3}{4})$, occupying octahedral interstitial sites. Inversion symmetry relates the two S atoms, while each Ni site lies on an inversion center. The structure consists of edge-sharing NiS$_6$ octahedra stacked along $c$, generating distinct in-plane and out-of-plane bonding geometries. At the first-order magnetostructural transition temperature $T_t \approx 265$~K, NiS undergoes an abrupt volume change of approximately $2\%$, with $a$ increasing by $\sim0.2\%$ and $c$ by nearly $1\%$~\cite{sparks2}. Importantly, the crystallographic space group $P6_3/mmc$ is preserved across this transition. Thus, the structural instability occurs within the same symmetry manifold, without symmetry lowering. In the low-temperature A-type AFM phase, Ni moments align ferromagnetically within each $ab$ plane and stack antiferromagnetically along the $c$ axis (Fig.~\ref{SI-fig1}(b)). This magnetic structure preserves the crystallographic lattice but breaks time-reversal symmetry $\mathcal{T}$. Inversion symmetry $\mathcal{P}$ remains intact, and the magnetic configuration is invariant under the combined antiunitary operation $\mathcal{S}=\mathcal{T}\tau_{1/2}$, where $\tau_{1/2}$ denotes translation by $c/2$ connecting the two Ni sublattices. Thus, while the high-temperature PM phase respects $P6_3/mmc1'$ (including $\mathcal{T}$), the low-temperature AFM phase retains inversion and non-symmorphic symmetry operations but belongs to a magnetic subgroup consistent with collinear order along $c$. The preservation of $\mathcal{P}$ ensures well-defined parity eigenvalues in the electronic structure, whereas the breaking of global $\mathcal{T}$ (and absence of $\mathcal{PT}$ symmetry) allows finite Berry curvature and symmetry-permitted Hall responses discussed in the main text.
\par 
Our calculations find three symmetry-distinct Ni–Ni exchange pathways which play crucial role in magnetism as schematically shown in Fig.~\ref{SI-fig1}(b). The first nearest-neighbor interaction ($J_1$) connects Ni atoms along the $c$ axis at a distance of approximately $2.70$~\AA. The midpoint of this bond preserves inversion symmetry, which forbids any Dzyaloshinskii–Moriya (DM) interaction for this exchange path. The second nearest-neighbor coupling ($J_2$) lies within the basal $ab$ plane at a separation of about $3.44$~\AA. Here as well, inversion symmetry at the bond midpoint suppresses antisymmetric exchange terms. In contrast, the third nearest-neighbor interaction ($J_3$), corresponding to a diagonal in-plane Ni–Ni separation of around $4.37$~\AA, lacks an inversion center at its bond midpoint. According to Moriya’s symmetry rules, this absence of inversion symmetry permits a finite DM vector for the 3NN exchange path as discussed in main text. The Ni–S–Ni bond angles strongly influence the superexchange interactions (see Fig.~\ref{SI-fig1}(c) and Table~\ref{struct}). The near-$90^\circ$ geometry of the 2NN pathway favors weak ferromagnetic exchange, while the larger deviations from $90^\circ$ for the 1NN and 3NN paths enhance antiferromagnetic $e_g$–$e_g$ superexchange.  Therefore, the combination of space-group symmetry (No.~194), non-symmorphic operations, and the A-type AFM configuration provides the structural and symmetry foundation for the exchange hierarchy, anisotropic interactions, and Berry-curvature–driven electronic responses analyzed in the main text.
\begin{table}[h!]
\centering
\caption{Experimental lattice parameters, Ni–Ni bond distances, and Ni–S–Ni bond angles of NiS at 77 K.}
\renewcommand{\arraystretch}{1.5}
\setlength{\tabcolsep}{15pt} 
\begin{tabular}{lc}
\toprule
\addlinespace[2pt]
\multicolumn{2}{l}{\textbf{Lattice Parameters (\AA)}} \\
\midrule
$a = b$ & 3.4456 \\
$c$     & 5.4051 \\
\midrule
\addlinespace[2pt]
\multicolumn{2}{l}{\textbf{Ni--Ni Bond Distances (\AA)}} \\
\midrule
$1^{st}$ NN & 2.70 \\
$2^{nd}$ NN & 3.44 \\
$3^{rd}$ NN & 4.37 \\
\midrule
\addlinespace[2pt]
\multicolumn{2}{l}{\textbf{Ni--S--Ni Bond Angles (\degree)}} \\
\midrule
$\theta_1$ & 68.4 \\
$\theta_2$ & 91.3 \\
$\theta_3$ & 132.4\\
\bottomrule
\end{tabular}
\label{struct}
\end{table}
\par 
The Fig.~\ref{SI-fig1}(d) shows the hexagonal Brillouin zone of NiS together with the momentum-space paths and planes used to characterize the altermagnetic electronic structure discussed in the main text. The spin-resolved band dispersions were calculated along the $L$–$\Gamma$–$\bar{L}$ (green dashed line) and $M$–$A$–$\bar{M}$ (purple dashed line) directions, which probe the momentum-dependent spin splitting characteristic of the altermagnetic state. In addition, the shaded horizontal ($k_x$–$k_y$) and vertical ($k_x$–$k_z$) planes passing through the vicinity of the $A$ point were used to compute the spin-resolved constant-energy contours at $E-E_F=-0.25$ eV as discussed in the main text. These contours reveal the anisotropic momentum-space spin polarization which is a hallmark of altermagnetism in NiS.
\section{Crystal Field Splitting and Electronic Structure}
To elucidate the microscopic origin of the band crossings discussed in the main text, we analyze the crystal-field splitting and orbital character of the Ni-$3d$ states using a Wannier-based tight-binding approach. We first construct a low-energy tight-binding model based on maximally localized Wannier functions (MLWFs), focusing on the non-spin-polarized electronic structure. As shown in Fig.~\ref{SI-fig2}(a), the Wannier-interpolated bands (red dashed lines) accurately reproduce the first-principles DFT dispersion (blue solid lines) along the high-symmetry directions of the hexagonal Brillouin zone. The excellent agreement validates the fidelity of the tight-binding representation. From the Wannier Hamiltonian, we extract the onsite block corresponding to the Ni-$3d$ manifold. In the absence of a crystalline environment, the five $d$ orbitals of a free Ni$^{2+}$ ion ($3d^8$ configuration) are degenerate. In the NiAs-type structure, however, each Ni atom is coordinated by six S ligands forming a slightly distorted octahedron. The resulting crystal-field potential lifts the fivefold degeneracy according to the local symmetry of the Ni site. Under the trigonal distortion compatible with space group $P6_3/mmc$, the cubic $t_{2g}$ and $e_g$ levels further split into irreducible representations of the local point group. As illustrated schematically in Fig.~\ref{SI-fig2}(b), the five Ni-$3d$ orbitals separate into three energy levels: a doubly degenerate $e_{1g}$ manifold (primarily $d_{xz}$ and $d_{yz}$), a doubly degenerate $e_{2g}$ manifold ($d_{x^2-y^2}$ and $d_{xy}$), and a non-degenerate $a_{1g}$ state derived from the $d_{z^2}$ orbital. From the onsite Hamiltonian extracted via Wannier interpolation, we obtain a crystal-field splitting of approximately $204$~meV between the $a_{1g}$ and $e_{2g}$ levels, and about $61$~meV between the $e_{2g}$ and $e_{1g}$ states. The relatively larger separation between $e_{2g}$ and $a_{1g}$ reflects the effect of trigonal distortion along the $c$ axis, consistent with the layered NiAs geometry. These energy scales provide the microscopic basis for the orbital hierarchy relevant to the low-energy electronic structure. Importantly, this crystal-field arrangement directly influences the symmetry character of the bands near the Fermi level. As discussed in the main text, the Dirac-like crossings near high-symmetry points arise from the interplay between bands carrying distinct orbital characters, primarily $e_{1g}$ and $e_{2g}$ symmetry. The extracted onsite splitting therefore provides a quantitative foundation for understanding the proximity and hybridization of these states.
\par 
We next analyze the total density of states (DOS) and partial DOS of Ni in the magnetically ordered state using spin-polarized LSDA+$U$+SOC calculations in the A-type antiferromagnetic configuration. The total density of states (inset of Fig.~\ref{SI-fig2}(c)) confirms the antiferromagnetic semimetallic character of NiS, characterized by a finite but low density of states at the Fermi level, in agreement with Ref.~\cite{mypaper}. Further PDOS (Figure~\ref{SI-fig2}(c)) shows the orbital-projected density of states (PDOS), resolved into $e_{1g}$, $e_{2g}$, and $a_{1g}$ contributions. The conduction band states near the Fermi level are predominantly composed of $e_{1g}$ character, with a smaller admixture of $e_{2g}$. In contrast, states just below the Fermi level contain significant $e_{2g}$ weight. This orbital distribution is fully consistent with the onsite crystal-field splitting extracted from the Wannier Hamiltonian. The proximity of $e_{1g}$- and $e_{2g}$-derived bands near $E_F$ naturally favors symmetry-enforced band crossings and enhances interband mixing once spin–orbit coupling is included. The combination of such crystal-field splitting, and he symmetry constraints imposed by the $P6_3/mmc$ lattice provides the essential ingredients for the Dirac-like band crossings analyzed in the main text. 
\bibliography{NiS}
\end{document}